\documentclass[onecolumn,authoryear]{els-mrw} 

\usepackage{amsmath,amssymb,amsfonts,amsthm,makeidx,graphicx}
\usepackage{txfonts}
\usepackage{helvet}
\usepackage{hyperref}

\begin{document}

\chapter{The first stars}\label{chap1}

\author[1]{Simon C.~O.\ Glover}%
\author[1,2]{Ralf S.\ Klessen}

\address[1]{\orgname{Universit\"{a}t Heidelberg}, \orgdiv{Zentrum f\"{u}r Astronomie, Institut f\"{u}r Theoretische Astrophysik}, \orgaddress{
Albert-Ueberle-Str.\ 2, 69120 Heidelberg, Germany}} 
\address[2]{\orgname{Universit\"{a}t Heidelberg}, \orgdiv{Interdisziplin\"{a}res Zentrum f\"{u}r Wissenschaftliches Rechnen}, \orgaddress{Im Neuenheimer Feld 225, 69120 Heidelberg, Germany}}

\articletag{Chapter Article tagline: update of previous edition,, reprint..}

\maketitle

\begin{glossary}[Glossary]

\term{Critical density} the density at which collisions with other particles start to dominate the de-excitation of excited states of atoms or molecules. 

\term{Collision-induced emission} cooling due to dipole transitions occurring during atom-molecule or molecule-molecule collisions. Unimportant at low densities owing to the very short collision timescale, but dominates the cooling of primordial gas above $n \sim 10^{14} \: {\rm cm^{-3}}$.

\term{Core-collapse supernova} a supernova explosion occurring at the end of the lifetime of a massive star due to the gravitational collapse of its nickel-iron core.

\term{Dark matter} hypothesised form of non-baryonic matter that interacts with normal matter only via gravity and (possibly) the weak nuclear force. Dark matter makes up the majority of the matter content of the Universe. Its true nature and properties are unknown.

\term{Dark matter halo} a quasi-spherical concentration of dark matter with a mean density roughly 200 times the mean cosmic matter density, supported against gravitational collapse by its own internal motions.

\term{Dark star} a hypothetical class of star supported by dark matter annihilation rather than by nuclear fusion.

\term{Dynamo} processes amplify preexisting magnetic fields in conducting media by converting kinetic energy into magnetic energy.

\term{Extremely metal poor star} a star with a metallicity (as measured by the ratio of Fe to H compared to the Solar value) that is less than $10^{-3}$ times the Solar metallicity.

\term{Free-fall collapse} refers to the rapid contraction of a body in which no other forces oppose gravitational attraction.

\term{Gravitational waves} are perturbations in the curvature of spacetime generated by the dynamic acceleration of massive objects, such as coalescing black holes or neutron stars, propagating as transverse waves at the speed of light.

\term{Hypernova} an unusually energetic form of core-collapse supernova, with an energy ten times larger than a standard supernova.

\term{Initial mass function} the initial distribution of masses of a population of stars. Because stars with different masses have very different lifetimes, the initial mass function can differ substantially from the present-day mass function.

\term{Jeans length} the critical length scale for gravitational instability.

\term{Jeans mass} the critical mass scale for gravitational instability, first derived by James Jeans in 1902.

\term{Lyman-Werner photons} ultraviolet photons with energies above 11.2~eV that are capable of photodissociating H$_{2}$.

\term{Metals} in astronomical terminology, any elements heavier than H and He.

\term{Pair instability supernova} an extremely energetic type of supernova predicted to occur when a very massive metal-free or low metallicity star becomes unstable because of the runaway production of electron-positron pairs.

\term{Population III} the first generation of metal-free stars

\term{Protostellar accretion disk} a rotationally-supported gas disk surrounding a protostar.

\term{Recovery time} the time delay between the explosion of the first Pop.\ III supernova in a dark matter halo and the formation of the first metal-enriched Population II stars in that halo.

\term{Self-shielding} the shielding of molecules against photodissociation by other molecules of the same type.

\term{Silk damping} the suppression of small-scale baryonic density perturbations in the early Universe owing to photon diffusion.

\term{Stellar archaeology} the use of elemental abundances measured in stars at the present day to learn about the earlier generations of stars responsible for producing these elements.

\term{Stellar feedback} the input by stars of energy and momentum into the gas surrounding them. Stellar feedback can in principle be positive (i.e.\ promoting the formation of additional stars) or negative (i.e.\ suppressing the formation of additional stars), but typically negative feedback dominates. 

\term{Streaming motion} a high redshift velocity offset between baryons and dark matter that is coherent on large scales.

\term{Yield} the mass of metals (or of a particular element) produced by a supernova.

\end{glossary}

\begin{glossary}[Nomenclature]
\begin{tabular}{@{}lp{34pc}@{}}
CCSN & Core-collapse supernova \\
CEMP & Carbon-enhanced extremely metal poor star \\
CIE & Collision-induced emission \\
DECIGO & DECi-hertz Interferometer Gravitational Wave Observatory (\href{http://tamago.mtk.nao.ac.jp/spacetime/decigo_e.html}{http://tamago.mtk.nao.ac.jp/spacetime/decigo\_e.html})\\
ET & Einstein Telescope  (\href{https://www.et-gw.eu/}{https://www.et-gw.eu/})\\
EMP & Extremely metal poor star \\
GW & Gravitational waves\\
HRD & Hertzsprung-Russell diagram \\
HST & Hubble Space Telescope (\href{https://science.nasa.gov/mission/hubble/}{https://science.nasa.gov/mission/hubble/}) \\
IGM & Intergalactic medium \\
IMF & Initial mass function \\
$\Lambda$CDM & The standard cosmological model that includes both a cosmological constant ($\Lambda$) and cold dark matter (CDM) \\
JWST & James Webb Space Telescope (\href{https://science.nasa.gov/mission/webb/}{https://science.nasa.gov/mission/webb/})\\
KAGRA & Kamioka Gravitational Wave Detector (\href{https://www.nao.ac.jp/en/research/telescope/kagra.html}{https://www.nao.ac.jp/en/research/telescope/kagra.html})\\
LIGO & Laser Interferometer Gravitational-wave Observatory (\href{https://www.ligo.caltech.edu/}{https://www.ligo.caltech.edu/})\\
LISA & Laser Interferometer Space Antenna (\href{https://lisa.nasa.gov/}{https://lisa.nasa.gov/})\\
LW & Lyman-Werner \\
MS & Main sequence \\
PDR & Photodissociation region \\
PISN & Pair instability supernova \\
Virgo & GW detector, hosted by the European Gravitational Observatory (\href{https://www.virgo-gw.eu/}{https://www.virgo-gw.eu/})\\
\end{tabular}
\end{glossary}

\begin{abstract}[Abstract]
Population III (or Pop.\ III) stars, the first stellar generation built up from metal-free primordial gas, first started to form at redshifts $z \sim 30$. They formed primarily in small dark matter halos with masses of a few million solar masses. The cooling of the gas in these halos was dominated on all scales by molecular hydrogen. Current theoretical models indicate that Pop.\ III stars typically formed in small clusters with a logarithmically flat mass function due to wide-spread fragmentation in the protostellar accretion disks around these primordial stars.  Massive Pop.\ III stars are thought to have played a pivotal role in shaping the early Universe, as their feedback regulates subsequent star formation, although the immediate effects of this feedback remain uncertain. Direct detection of Pop.\ III stars is challenging, but our chances of detecting at least a few Pop.\ III supernovae within the next decade are brighter. Indirect approaches based on stellar archaeology or gravitational wave detections offer promising constraints. Current observations suggest that most massive Pop.\ III stars ended their lives as core-collapse supernovae rather than pair-instability supernovae, offering insight into the initial mass function and evolutionary pathways of these primordial stars.
\end{abstract}

\begin{BoxTypeA}[box]{Key points}
\begin{itemize}
\item The first Population III stars formed at redshifts $z \sim 30$ in dark matter halos with masses of around $10^{6} \: {\rm M_{\odot}}$.
\item The thermal evolution of Population III star-forming gas in these halos is governed on almost all scales by molecular hydrogen (H$_{2}$). The resulting gas temperatures are significantly higher than what is typical for present-day star formation.

\item The formation of a Pop.\ III protostar is rapidly followed by the build-up of a gravitationally unstable protostellar accretion disk. This disk is highly susceptible to fragmentation, resulting 
in the formation of a dense cluster of Pop.\ III protostars. These protostars have a wide range of masses, with the overall distribution being much flatter than the present day stellar initial mass function.

\item Feedback from Pop.\ III stars plays an important role in quenching star formation within the earliest star-forming halos. On small scales, photoionisation dominates, while on much larger scales,  Lyman-Werner feedback plays a key role in suppressing subsequent Pop.\ III star formation.

\item Although we are unlikely to be able to observe massive Pop.\ III stars at any point in the near future, direct detection of the supernova explosions occurring at the end of their lives is a more promising avenue. In addition, stellar archaeology and the detection of gravitational waves from merging black holes provide important constraints on the Pop.\ III IMF.
\end{itemize}
\end{BoxTypeA}

\section{Introduction}\label{intro}
The formation of the first stars in the Universe -- the so-called {\bf Population III} (in short Pop.\ III) stars\footnote{In Baade's original classification of stellar populations in spiral galaxies, he identified two main populations: young, high metallicity stars, which he termed Population I and older, lower metallicity stars, which he called Population II. The idea that there must have been a Population III, an early generation of stars responsible for producing the heavy elements observed in the spectra of Population II stars, was a much later introduction. Originally, this term was used to refer to any early stars that formed with metallicities lower than around $10^{-3} \: {\rm Z_{\odot}}$ \citep[see e.g.][]{bond81}, but over time it has come to refer exclusively to stars forming from completely metal-free gas.} -- is a key milestone in the history of our Universe. At the earliest cosmic epochs, the matter distribution in our Universe was extremely smooth. Over time, structure developed due to the influence of gravity, and at a time of around 100~Myr after the Big Bang, gas became able to gravitationally collapse to densities high enough to enable the formation of the very first stars. Feedback from these stars profoundly affected their surroundings, and so understanding their birth, evolution and death is crucial for understanding the earliest stages of galaxy formation and evolution. In this chapter, we briefly summarise our current state of knowledge regarding Pop.\ III. In Section~\ref{sec:where}, we discuss the factors that determine when and where Pop.\ III star formation begins. In Section~\ref{sec:how}, we introduce the physics involved in the gravitational collapse of metal-free gas from intergalactic to protostellar densities, and we also discuss what we currently understand regarding the {\bf initial mass function} (IMF) of Pop.\ III stars. We provide an overview of the most important forms of feedback produced by Pop.\ III stars in Section~\ref{sec:feedback}, and in Section~\ref{sec:obs} we discuss the prospects for studying Pop.\ III star formation observationally. We close with a brief summary in Section~\ref{summary}.

\section{Where and when do the first stars form?}\label{sec:where}

In the conventional cold dark matter paradigm with cosmological constant,  the $\Lambda$CDM model, structure in the Universe develops in a hierarchical fashion. Starting from an initially almost smooth state, small perturbations grow in the dark matter component on all scales due to gravitational instability. The growth of these perturbations depends on their overdensity $\delta \equiv \rho / \bar{\rho}_{\rm m} - 1$, where $\rho$ is the density of the perturbation and $\bar{\rho}_{\rm m}$ is the mean matter density in the Universe. While $\delta \ll 1$, the growth of the perturbations is accurately described by linear perturbation theory, which predicts that at high redshifts $\delta \propto (1 + z)^{-1}$. However, once the overdensity becomes large ($\delta \sim 1$), the evolution of the perturbations becomes non-linear and their growth accelerates. Simple analytical considerations and N-body simulations both demonstrate that the outcome of this non-linear evolution is a filamentary web of moderately overdense dark matter structures (the ``Cosmic Web''), with highly overdense, quasi-spherical concentrations of dark matter located at the junctions of this web. These highly dense regions are known as {\bf dark matter halos}, and they are the locations where stars and ultimately galaxies will eventually form. In $\Lambda$CDM, the first dark matter halos to form have very low masses\footnote{The minimum mass scale depends on the model adopted for the dark matter, but for example is of the order of an Earth mass or smaller in models in which dark matter is a weakly interacting massive particle
\citep[see e.g.][]{diemand05}.}, with larger halos forming at later times through a mix of mergers and accretion. 

In contrast to the dark matter, the growth of structure in the baryonic component of the Universe (henceforth referred to simply as ``gas'') is complicated by the effects of gas pressure. This suppresses gravitational instability on scales smaller than a critical length scale known as the {\bf Jeans length}:
\begin{equation}
L_{\rm J} \equiv c_{\rm s} \sqrt{\frac{\pi}{G \rho}},
\end{equation}
where $c_{\rm s}$ is the sound speed. We can also define a critical mass scale associated with $L_{\rm J}$, the {\bf Jeans mass}:
\begin{equation}
M_{\rm J} = \frac{4\pi}{3} \rho \left( \frac{L_{\rm J}}{2} \right)^{3}.  
\end{equation}
Evaluating this, we find that at densities close to the mean cosmic density, $M_{\rm J} \simeq 10^{5} \: {\rm M_{\odot}}$ at redshifts $z > 100$ and $M_{\rm J} \simeq 10^{5} [(1+z) / 100]^{3/2} \: {\rm M_{\odot}}$ at lower redshifts.
At early times, when the characteristic dark matter halo mass scale is much smaller than $M_{\rm J}$, the gas distribution remains relatively smooth. Highly overdense concentrations of gas only develop once dark matter halos with masses $M > M_{\rm J}$ assemble. The precise time at which this occurs depends on the cosmological model, but in the standard $\Lambda$CDM model, halos with $M > M_{\rm J}$ start to become reasonably common at redshifts of around $z \sim 30 -40$. 

The assembly of dark matter halos with masses above $M_{\rm J}$ is a necessary condition for the onset of Pop.\ III star formation, but not a sufficient condition. As gas falls into the potential well associated with the dark matter halo, it heats up, owing to a combination of adiabatic compression and shock heating. If the gas is unable to radiate away this thermal energy, then its Jeans mass increases until the system once again becomes pressure supported. Therefore, a second necessary condition for the onset of Pop.\ III star formation is that the gas be able to radiate away a substantial fraction of the thermal energy it gains during its collapse within a timescale that is substantially shorter than a Hubble time. Gas in these first cosmic structures is chemically simple: it has an elemental composition set by the outcome of primordial nucleosynthesis, and hence consists primarily of hydrogen and helium, with trace amounts of deuterium and lithium. None of these elements have fine structure transitions in their ground state, and so atomic cooling becomes possible only at temperatures high enough to excite permitted dipole transitions, such as the hydrogen Lyman-$\alpha$ line. In practice, this corresponds to a minimum gas temperature $T \sim 10^{4} \: {\rm K}$, a value that is around 10--40 times larger than the temperatures actually found in these first structures. In these very earliest structures, atomic cooling is therefore completely ineffective, and cooling from molecules -- specifically, molecular hydrogen (H$_{2}$) dominates.

In the local interstellar medium, H$_{2}$ forms primarily on the surface of dust grains. However, the formation of dust requires elements heavier than H and He, which are not present in the primordial gas, and hence grain surface formation of H$_{2}$ is impossible there. Instead, H$_{2}$ is formed primarily via the reaction chain 
\begin{align} 
{\rm H + e^{-}} & \rightarrow  {\rm H^{-} + \gamma}, \\
{\rm H^{-} + H} & \rightarrow  {\rm H_{2} + e^{-}},   
\end{align}
or by a similar but less important chain involving H$^{+}$ and H$_{2}^{+}$. Formation of H$_{2}$ via this reaction chain is strongly suppressed at $z \gg 100$ by photodetachment of the H$^{-}$ ions by the cosmic microwave background, but this becomes unimportant at lower redshifts, allowing a small amount of H$_2$ to form in the intergalactic gas prior to the formation of structure. Detailed calculations show that the resulting H$_{2}$ fraction is $x_{\rm H_{2}} \sim 10^{-6}$ \citep{hirata06}, with the build-up of additional H$_{2}$ being limited primarily by the time it takes the molecule to form. This tiny H$_{2}$ fraction is too small to provide effective cooling of the gas in the first structures, and so the question of whether or not this gas can cool ultimately depends on the amount of H$_2$ that forms during the infall of gas into the halo.

Modelling of the chemical and thermal evolution of gas in the first structures 
\citep[see e.g.][]{tegmark97,yoshida03,schauer2019} shows that halos with masses just above $M_{\rm J}$ never form enough H$_{2}$ to cool effectively. However, these models also demonstrate that the amount of H$_{2}$ formed increases with increasing halo mass, while at the same time the amount of H$_{2}$ required decreases. In both cases, this behaviour is driven by the gas temperature. As the halo mass increases, infalling gas is heated more strongly and reaches a higher peak temperature. This increased temperature leads to increased formation of H$^{-}$ and hence H$_{2}$, and also enables a larger number of the rotational and vibrational energy levels of H$_{2}$ to be excited, leading to faster radiative cooling. In practice, models in which the gas starts at rest with respect to the dark matter and evolves quiescently (i.e.\ without significant dynamical heating from mergers or rapid gas infall) typically find that a peak temperature of approximately $T_{\rm peak} \sim 1000$~K is required to enable efficient cooling, corresponding to a critical mass scale
\begin{equation}
M_{\rm crit} \simeq 3 \times 10^{5} \left(\frac{1+z}{30} \right)^{-3/2} \: {\rm M_{\odot}},
\label{eq:mcrit}
\end{equation}
around an order of magnitude larger than $M_{\rm J}$. Accounting for dynamical heating, e.g.\ from rapid accretion or mergers with other halos, increases this number by a further factor of a few \citep{schauer2019}.

\begin{figure}
\vspace{-0.5cm}
\includegraphics[width=0.66\textwidth]{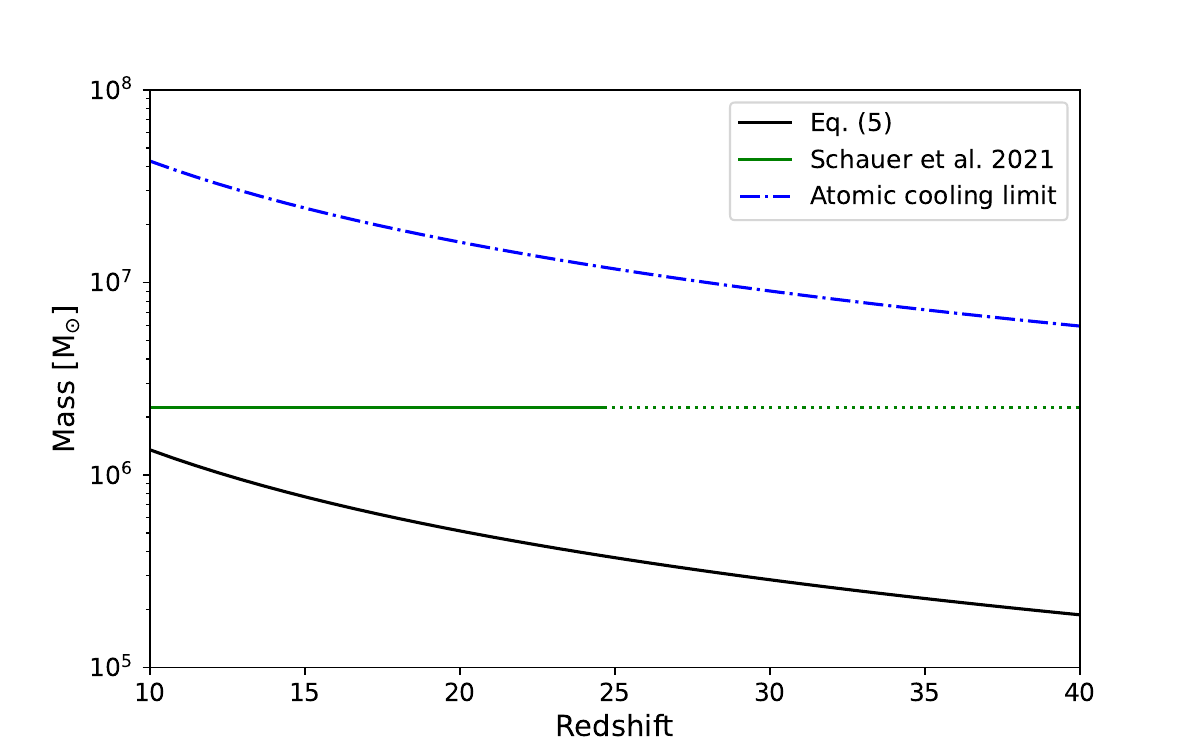}
\hspace{-0.3cm}
\begin{minipage}[b]{0.30\textwidth}
\caption{Comparison of predictions for the critical halo mass scale required in order to form Pop.\ III stars, $M_{\rm crit}$. The black line shows the redshift evolution predicted by models that do not account for dynamical heating or baryonic streaming (see Eq.~\ref{eq:mcrit}), while the green line is a fit to cosmological simulation results from \citet{schauer21} that account for both effects. Note that this fit is based only on data for redshifts $z < 25$; our extrapolation to higher redshifts is indicated with a dotted line. For comparison, we also show the mass scale above which Lyman-$\alpha$ cooling is expected to dominate (blue dot-dashed line). 
\vspace*{0.5cm}}
\label{fig:mcrit}
\end{minipage}
\end{figure}

A final complication comes from the fact that in general the gas does not start at rest with respect to the dark matter. At very early epochs the baryonic and radiation components of the Universe are strongly coupled to one another and evolve as a coupled baryon-photon fluid. Small-scale perturbations in this coupled fluid are suppressed by photon diffusion (also known as {\bf Silk damping}), and so only large-scale perturbations to the density and velocity fields survive. Since the same is not true for the dark matter, we therefore expect the baryon-photon fluid to move relative to the dark matter. After the gas and radiation decouple at $z \sim 1000$, the gas can start respond to the small-scale fluctuations in the dark matter, and so given enough time, the relative motion between the two components would be lost. However, this does not happen instantaneously. Instead, the baryons retain a motion relative to the dark matter, with a root-mean-squared (rms) amplitude of $\sim 30 \: {\rm km \: s^{-1}}$ at $z \sim 1000$, decreasing as $(1+z)$ at lower redshifts \citep{tsel2010}. The coherence length of this relative motion is orders of magnitude larger than the sizes of the first halos, and so on the small scales relevant for their evolution, we effectively have a uniform velocity offset (or {\bf streaming motion}) between the gas and the dark matter. This streaming motion acts to displace the centre of collapse of the gas from the centre of the dark matter potential well. The importance of this depends on the strength of the streaming velocity. In the common case ($v_{\rm stream} \sim v_{\rm rms}$), it has a moderate impact on the gas density, reducing its ability to form H$_{2}$, increasing the amount of H$_{2}$ required for effective cooling, and consequently increasing $M_{\rm crit}$ by a factor of a few compared to the no streaming case (see e.g.\ \citealt{fialkov12,schauer2019} or the comparison in Figure~\ref{fig:mcrit}). On the other hand, in the rarer patches of the Universe in which $v_{\rm stream}$ is significantly larger than $ v_{\rm rms}$, streaming has a much stronger effect, strongly suppressing H$_{2}$ formation and cooling in all halos with peak temperatures $T < 10^{4}$~K \citep{schauer2019}.

To summarise: the first dark matter halos massive enough to produce large overdensities in the gas have masses of a few times $10^{4} \: {\rm M_{\odot}}$. However, H$_{2}$ cooling is not effective in these small halos, and so the gas associated with them is unable to undergo prolonged gravitational collapse and hence is prevented from forming stars. The first halos in which H$_{2}$ cooling becomes efficient enough to enable Pop.\ III stars to form have masses of between $3 \times 10^{5} \: {\rm M_{\odot}}$ (in the extreme case of negligible dynamical heating and no streaming) to $10^{6} \: {\rm M_{\odot}}$ (in the more typical case of moderate dynamical heating and average streaming). Halos with these masses start to form in significant numbers at a redshift of around 30 (corresponding to a time of around 100~Myr after the Big Bang), and so it is this moment that for practical purposes marks the onset of Pop.\ III star formation.

\section{How do the first stars form?}\label{sec:how}
\subsection{The initial collapse phase}
\label{sec:init_coll}
The mean number density of the gas in the first halos, prior to the onset of H$_{2}$ cooling, is around $n \sim 1 \: {\rm cm^{-3}}$. For comparison, the mean number density of the plasma making up a main sequence star is $n \sim 10^{24} \: {\rm cm^{-3}}$. In order for gas to move from halo densities to stellar densities, it therefore needs to increase its number density by 24 orders of magnitude, decreasing its characteristic size scale by $\sim 8$ orders of magnitude in the process. This enormous change in density is associated with the release of a large amount of gravitational potential energy, the majority of which must be radiated away if gravitational collapse is to continue. Therefore, in order to understand how primordial gas gets from halo densities to stellar densities, we need to study how it cools. 

As we have already discussed, cooling in primordial gas at the temperatures relevant for Pop.\ III star formation is dominated by rotational and vibrational emission from H$_{2}$, with rotational emission dominating at temperatures $T < 1000$~K and vibrational emission becoming important at higher temperatures. This has two important consequences. First, because H$_{2}$ is a very light molecule with a low moment of inertia, there are large energy separations between its rotational energy levels: for example, the $J=0$ and $J=1$ levels are separated by $\Delta E_{10}/k \simeq 128$~K, while the $J = 0$ and $J = 2$ levels are separated by $\Delta E_{20}/k \simeq 512$~K, where $k$ is the Boltzmann constant. In addition, transitions between levels with odd $J$ and even $J$ involve a change in nuclear spin and are highly forbidden, so the main contribution to the cooling rate at low temperatures comes from quadrupole transitions from $J = 2$ to $J = 0$. As a result, the H$_{2}$ cooling rate falls off exponentially with decreasing temperature for $T < 500$~K. The minimum temperature that can be reached with H$_{2}$ cooling alone depends to some extent on the gas density and the fractional abundance of H$_{2}$, but is usually around 150--200~K. Below this temperature, the exponential fall-off in the cooling rate renders H$_2$ cooling completely ineffective. HD, the singly deuterated analogue of H$_{2}$, can in some circumstances provide effective cooling at lower temperatures, but this typically requires either highly quiescent collapse or the pre-processing of the gas by a previous period of ionisation \citep{mcgreer08} and it plays little role in the formation of the first generation of Pop.\ III stars.

Second, the radiative transition probabilities of the rotational and vibrational transitions of H$_{2}$ are relatively small, owing to the molecule's lack of a dipole moment, and it therefore has a low {\bf critical density}, $n_{\rm crit}$. This is the density at which collisions with other atoms and molecules become more important than radiative processes for determining the level populations; in the case of H$_{2}$, it has a value of approximately $n_{\rm crit, H_{2}} \sim 10^{4} \: {\rm cm^{-3}}$. The importance of the critical density in this context stems from the fact that at densities $n < n_{\rm crit}$, the cooling rate depends on both the number density of colliders and the number density of H$_{2}$ molecules and hence scales as $\Lambda_{\rm H_{2}} \propto n^{2}$, whereas at densities $n > n_{\rm crit}$, the cooling rate becomes insensitive to the number density of colliders and scales as $\Lambda_{\rm H_{2}} \propto n$. For comparison, gas undergoing gravitational collapse with minimal pressure support is heated by adiabatic compression at a rate that scales with density as $\Gamma_{\rm pdv} \propto n^{3/2}$. Therefore, at low densities cooling becomes more effective than heating as the density increases, whereas at high densities the opposite is true.

\begin{figure}[t]
\centering
\includegraphics[width=0.95\textwidth]{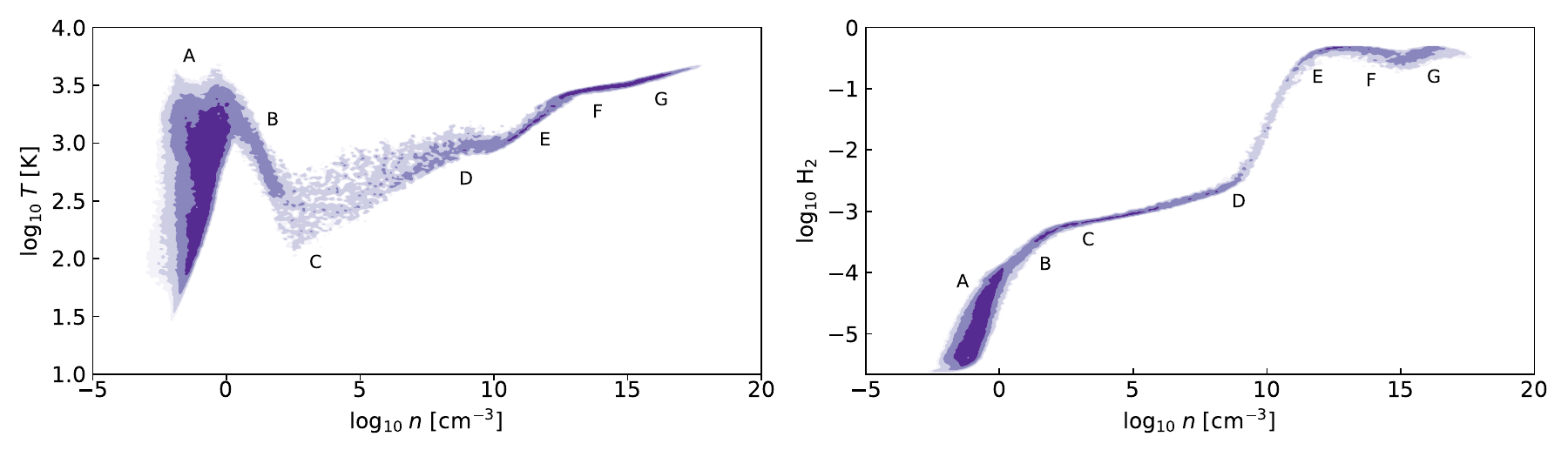}
\caption{Temperature $T$ as a function of H nuclei number density $n$ 
shortly before the formation of the first Pop.\ III protostar in a dark matter halo. The labels indicate various points of interest: (A) initial heating phase as gas falls into dark matter halo; (B) rapid cooling following H$_{2}$ formation; (C) H$_{2}$ cooling becomes much less effective as gas approaches $T_{\rm min} \sim 200$~K and $n$ exceeds $n_{\rm crit}$; (D) gradual heating once compressional heating dominates over radiative cooling; (E) three-body H$_{2}$ formation converts remaining atomic hydrogen to H$_2$, collapsing gas becomes fully molecular; (F) onset of cooling due to collision-induced emission; (G) gas becomes optically thick, further cooling possible only due to H$_{2}$ collisional dissociation. Modified version of a figure from \citet{kg23}, based on data supplied by L.\ Prole.}
\label{fig:collapse}
\end{figure}

The microphysics of H$_{2}$ cooling therefore introduces both a characteristic density and a characteristic temperature into the problem. Simulations of the gravitational collapse of primordial gas
\citep{abn02,bromm02,yoshida03} show that at $n < n_{\rm crit}$, the gas cools as it collapses, lowering its temperature from an initial value of a few thousand K to the minimum value of $\sim 200$~K reachable by H$_{2}$ cooling alone (see point C in Figure~\ref{fig:collapse}). During this phase of the collapse, the Jeans mass,  which scales with density and temperature as $M_{\rm J} \propto T^{3/2} n^{-1/2}$, decreases substantially, leading in many cases to the gas breaking up into smaller scale fragments with masses of order the local Jeans mass. This behaviour changes, however, once the density reaches $n_{\rm crit}$. Above that value, heating becomes more effective than cooling and the temperature of the gas begins to increase
(Figure~\ref{fig:collapse}, point D). Because of the steep temperature dependence of the H$_{2}$ cooling rate, this increase is relatively slow, scaling with density approximately as $T \propto n^{1/10}$. As a result, the Jeans mass continues to decrease during this phase of the collapse, albeit at a much slower rate. Fragmentation during this phase of the collapse is not very efficient -- the gas can still fragment, but it is much less likely to do so in the absence of large external perturbations. As a result, the characteristic mass of the collapsing clumps remains close to the value of the Jeans mass at $n_{\rm crit}$, i.e.\ a few hundred solar masses.

The next event of note occurs once the density reaches $n \sim 10^{9} \: {\rm cm^{-3}}$. At this point, the timescale to convert hydrogen from atomic to molecular form via the three-body reaction
\begin{equation}
\rm H + H + H \rightarrow H_2 + H,
\end{equation}
and analogous reactions involving H$_{2}$ and He as the third body, becomes shorter than the {\bf free-fall collapse} timescale. Therefore, once the gas reaches this density, wholesale conversion of H to H$_{2}$ occurs (Figure~\ref{fig:collapse}, point E). This chemical transition increases the abundance of the dominant coolant by a factor of a 1000, and naively one would expect this to have a profound impact on the thermal evolution of the gas. In practice, however, models show that the impact is actually very small, because the increased cooling is largely offset by two other effects: the strong chemical heating associated with the transition from H to H$_{2}$, and the fact that H$_{2}$ line emission becomes optically thick at around this density. 

As the collapse proceeds further, H$_{2}$ line cooling gradually becomes more and more ineffective, until at very high densities ($n > 10^{14} \: {\rm cm^{-3}}$; see Figure~\ref{fig:collapse}, point F), a new cooling mechanism starts to become important: {\bf collision-induced emission} (CIE). Although the H$_{2}$ molecule has no dipole moment, the reaction complex formed when H$_{2}$ collides with e.g.\ an H atom or another H$_{2}$ molecule, does have a dipole moment and can undergo dipole radiative transitions. Because the collision timescale is very short, the probability that any given collision leads to the emission of a photon during the collision is very small. However, the sheer number of collisions that occur as we move to very high densities means that this process eventually becomes the dominant cooling mechanism \citep{ra04}. Initially, the large linewidth associated with CIE cooling means that the optical depth is small. However, once the density exceeds $n \sim 10^{15} \: {\rm cm^{-3}}$, this process also becomes optically thick (Figure~\ref{fig:collapse}, point G). From this point on, radiative cooling is no longer effective at cooling the gas. It therefore starts to heat up adiabatically until it reaches a temperature of $T \sim 2000 \: {\rm K}$, at which point the H$_{2}$ molecules start to dissociate. In order to fully dissociate the H$_{2}$, we need 4.48~eV of energy per H$_{2}$ molecule, and so H$_{2}$ collisional dissociation acts as a kind of thermostat, maintaining the temperature at close to 2000~K until all of the H$_{2}$ is gone.\footnote{A useful analogy is with boiling water, which remains at 100 $^\circ$C even in the presence of ongoing heating until all of the water has been converted to steam.}

Once the H$_{2}$ is gone, the gas undergoes a second phase of adiabatic evolution until its temperature reaches $10^{4}$~K. At this point, the atomic hydrogen starts to ionise, again acting as an effective thermostat, maintaining the temperature at $T \sim 10^{4}$~K until the gas is almost fully ionised. Finally, once the atomic hydrogen is gone, the gas starts its final phase of adiabatic evolution, which terminates once the local Jeans mass increases to match the mass of the collapsing gas. At this point, a hydrostatically-supported core is formed \citep[cf. the similar behaviour for present-day star formation described in e.g.][]{larson69}. The mass of this core is set by the Jeans mass at the start of the final adiabatic phase and is typically a few times 0.01~M$_{\odot}$. The core radius is around 0.04 AU (equivalent to about 10~R$_{\odot}$). This hydrostatic core is the first object that we can identify as a true Pop.\ III protostar.

\subsection{Disk formation and fragmentation}
Although the initial mass of a Pop.\ III protostar is very small, it is embedded in a much larger clump of gas with a mass of a few hundred solar masses, most of which is falling in toward the protostar. The final mass of the protostar therefore depends upon what happens to this gas. There are three obvious scenarios:
\begin{enumerate}
\item The protostar accretes a large fraction of the clump mass. In this case, its final mass will be of a similar order as the clump mass,
i.e.\ a few hundred stellar masses.
\item The protostar accretes a small fraction of the clump mass, with the remainder of the gas being driven away from the protostar
by some form of stellar feedback. In this case, the final stellar mass depends on the efficiency of the feedback (i.e.\ on what fraction
of the clump mass it can remove).
\item The clump fragments and forms multiple protostars. This leads to the formation of a dense cluster with members competing for further accretion from the available gas reservoir. Some protostars may also get ejected prematurely by stellar dynamical effects. The end result is a collection of stars with a range of masses, leading to a wide Pop.\ III {\bf initial mass function} (IMF)
\end{enumerate}
In practice, scenario 2 can be quickly ruled out. The high temperature of the clump implies a very high mass inflow rate, typically $\dot{M} \sim 0.01 \: {\rm M_{\odot}} \: {\rm yr^{-1}}$. Therefore, the time required for a significant fraction of the clump mass to flow to the centre is only a few times $10^{4} \: {\rm yr}$. Efficient stellar feedback requires the Pop.\ III star to have reached the main sequence (see Section~\ref{sec:feedback} below) and to have started fusing hydrogen to helium, something which occurs $\sim 10^{4} \: {\rm yr}$ or more after the formation of the protostar \citep{op03}. Therefore, by the time that stellar feedback starts to act, a significant fraction of the clump mass will already have reached protostellar densities, ruling out models in which the majority of the clump mass is dispersed by feedback.

Early numerical studies of Pop.\ III star formation \citep[e.g.][]{abn02,bromm02} were for technical reasons largely restricted to following the initial collapse phase described in Section~\ref{sec:init_coll}. Very limited fragmentation of the gas occurs during this phase once $n > n_{\rm crit}$, and studies of Pop.\ III star formation often assumed that the same would hold true after the formation of the first protostar, i.e.\ that the outcome would resemble scenario 1. This was used to argue in favour of a picture in which all Pop.\ III stars were massive, with masses typically in excess of $100 \: {\rm M_{\odot}}$.

\begin{figure}
\vspace*{0.2cm}
\includegraphics[width=0.66\textwidth]{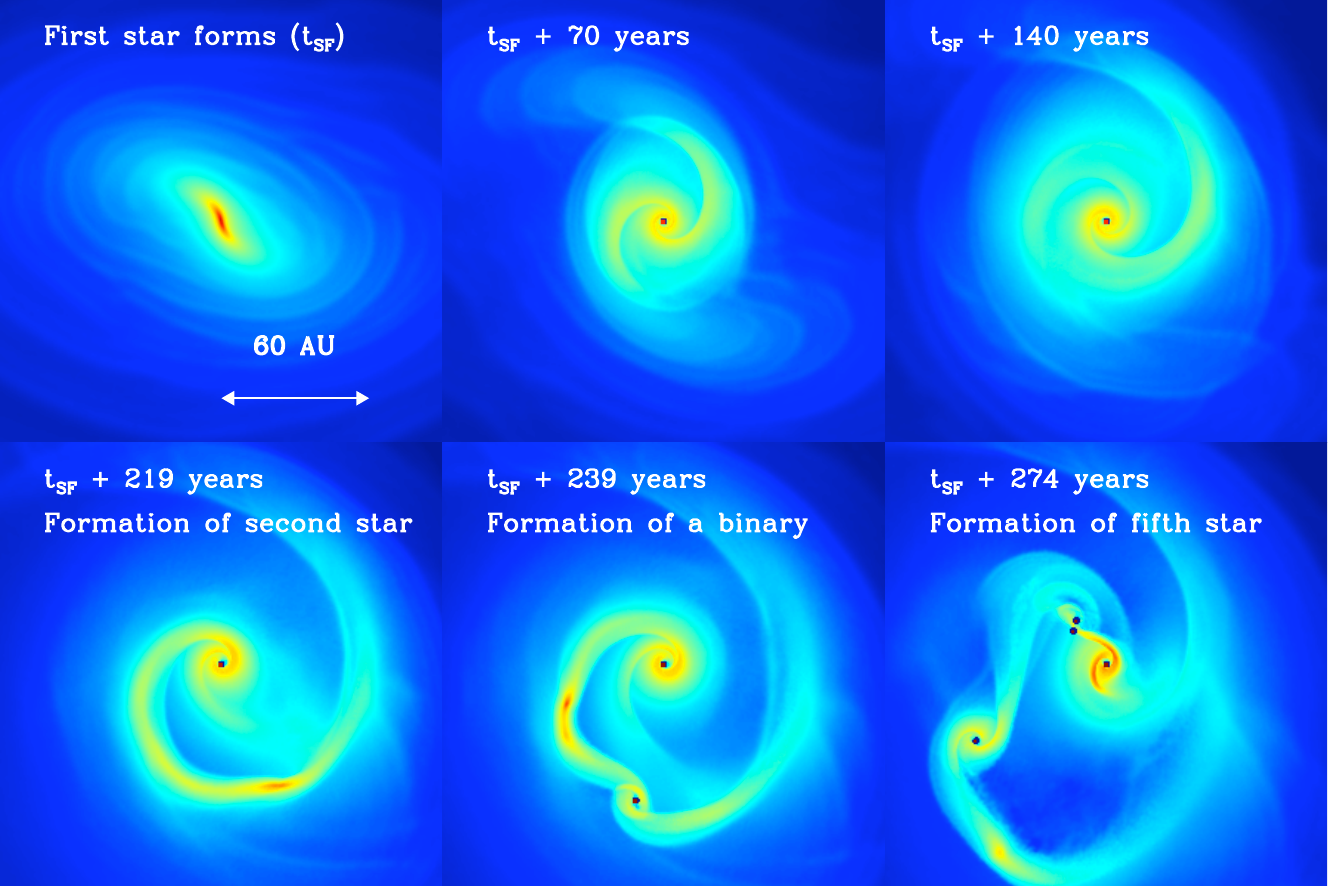}
\hspace{0.3cm}
\begin{minipage}[b]{0.30\textwidth}
\caption{Formation and evolution of the disk around a Pop.~III protostar in a simulation by \citet{Clark11}. It is clearly noticeable that the disk quickly becomes gravitationally unstable and develops a two-arm spiral structure which then fragments into four protostars within 240 years. Figure from \citet{Clark11}, reproduced with permission.}
\label{fig:disk}
\end{minipage}
\end{figure}

However, we now understand that this picture is incorrect. Because the inflowing gas typically has non-zero angular momentum, it is unable to flow directly onto the initial protostar. Instead, it forms a rotationally supported {\bf protostellar accretion disk} surrounding the star, in a similar fashion to what is observed to happen in present-day star formation \citep{McKee2007}. Turbulent viscosity within this disk allows angular momentum to be transferred outwards and gas to flow inwards, onto the protostar. Models of the dynamics of protostellar accretion disks show that there is an upper limit on the rate at which mass can flow inwards through the disk, set by the disk temperature and surface density \citep[see e.g.][]{tan04,Peters2010}. If the accretion rate onto the disk is sufficiently small, a steady state solution is possible in which the flow of mass onto the disk is balanced by the flow of mass from the disk onto the protostar. However, in the case of Pop.\ III star formation, the accretion rate onto the disk is too large for this to be possible. Instead, the disk steadily accumulates mass, increasing its surface density, until it eventually becomes gravitationally unstable. At this point, it develops pronounced spiral density waves which soon thereafter fragment, forming additional protostars, as illustrated in Figure~\ref{fig:disk}  \citep[see also e.g.][]{Clark11,greif11}.

The subsequent evolution of the protostellar system is complex and still not fully understood. The fragments formed in the disk can themselves act as centres of collapse, accreting gas from their surroundings. 
A significant number migrate inwards through the disk, ultimately merging with the central protostar \citep{chon19}, while others are ejected as a result of dynamical encounters between fragments, in some cases remaining gravitationally bound to the centre of mass of the system, and in other cases being ejected from the central region entirely \citep{prole22}. Efforts to numerically simulate the evolution of the system are hampered by the large range of timescales involved: the characteristic dynamical timescale on the scale of the individual fragments is less than a year, but even in models with stellar feedback, the lifetime of the disk is many tens of thousands of years. For this reason, simulations typically have to compromise either on spatial resolution (thereby missing much of the small-scale fragmentation) or on the total time followed (thereby limiting what we can say about the late-time evolution of the system). These difficulties mean that as yet, no fully satisfactory model exists for the IMF of Pop.\ III stars. That said, a common feature of essentially all current efforts is that they predict an IMF which is approximately logarithmically flat, i.e.\ one in which
\begin{equation}
\frac{dn}{d\log M} \simeq {\rm constant},
\end{equation}
where $n(M)$ is the number of stars of mass $M$. This form for the IMF is dramatically different from the IMF of present-day stars, which peaks at a mass of around 0.2--0.3~M$_{\odot}$ and drops rapidly as a power-law at higher masses \citep[][]{Kroupa2002}. An important consequence is that whereas most of the mass in a present-day stellar population is accounted for by stars close to the peak of the IMF, the bulk of the mass in a Pop.\ III stellar population is instead accounted for by the most massive stars. In this sense, the Pop.\ III IMF is top heavy compared to the present day IMF, a fact that is generally accepted within the field despite the uncertainties that remain regarding the precise slope of the IMF, as well as its low mass and high mass cutoffs.

\subsection{Additional processes affecting fragmentation}
Aside from stellar feedback, which we discuss in the next section, there are two additional physical processes that may play important roles in regulating fragmentation in Pop.\ III protostellar accretion disks. These are energy input due to dark matter annihilation, and the support provided by magnetic fields. We briefly introduce both processes  below and refer the reader to \citet{kg23} for a more in-depth discussion.

\subsubsection{Dark matter annihilation}
In some particle physics models for dark matter, the dark matter particle is a Majorana particle, i.e.\ it is its own anti-particle. There is therefore a non-zero probability that encounters between dark matter particles lead to annihilation, converting the rest mass energy of the particle pair into some combination of photons, neutrinos and stable charged particles. Typically, the annihilation cross-section is small and so the energy input from dark matter annihilation plays little role during the early stages of the gravitational collapse of the gas. However, this may be different at the high densities found at the centre of dark matter halos. For example, \citet{spolyar08} and \citet{freese08} posit that gravitational collapse will be halted once the energy input from dark matter annihilation exceeds the H$_{2}$ cooling rate and argue that this will lead to the formation of so-called {\bf dark stars} supported by annihilation rather than by nuclear fusion. If correct, this scenario would lead to a profound difference in the outcome of the Pop.\ III star formation process compared to the conventional scenario. However, \citet{smith12} studied the impact of dark matter annihilation using 3D hydrodynamical simulations and reported the gravitational collapse continues past the point where \citet{spolyar08} predict it to stop, aided by the strong temperature dependence of the H$_{2}$ cooling function and the chemical cooling provided by H$_{2}$ collisional dissociation and H ionisation. They found no convincing evidence for the formation of dark stars, but noted that the heating from dark matter annihilation played an important role in stabilising the protostellar accretion disk against fragmentation. We also point out that one common feature in these studies is the assumption that the central dark matter density in the minihalo is substantially boosted by adiabatic contraction as the gas collapses \citep[see e.g.][]{blumenthal86}. This is a reasonable approximation in a highly symmetrical, smooth collapse, but \citet{stacy12,stacy14}  showed that in more realistic conditions, the presence of non-axisymmetric perturbations rapidly brings about a separation between  the location of the collapsing gas and the dark matter density cusp (which is therefore not strongly boosted), resulting in dark matter annihilation having only a negligible impact on the evolution of the gas. 

\subsubsection{Magnetic fields}
Although magnetic fields can be produced early in the history of the Universe by a variety of physical mechanisms \citep[see e.g.][for an overview]{durrer13}, the tiny strength of the resulting fields was for a long time used to argue that they would play no role in Pop.\ III star formation. However, we now understand that even a minuscule seed field can be exponentially amplified by the turbulent {\bf dynamo} acting within the first star-forming dark matter halos, allowing the field strength to reach dynamically significant values in less than a free-fall time \citep[see e.g.][]{kulsrud97,Schleicher2010,schober12}. Nevertheless, the extent to which this field actually impacts Pop.\ III star formation remains unclear. 

Early work on Pop.\ III formation in magnetised gas found the field to have a large impact, suppressing fragmentation whenever the magnetic field energy exceeds the rotational energy, and powering a jet (\citealt{machida08}; see also the more recent simulations by \citealt{sadanari21}). However, this work assumed an initially ordered field, whereas the field generated by the turbulent dynamo is expected to be highly disordered on all scales. Simulations starting with an initially highly tangled field find more mixed results, with results ranging from complete \citep{hirano22} or partial \citep{sharda20,sadanari24} suppression of fragmentation to no significant impact \citep{prole22_mhd}. Resolving this disagreement remains a topic of active research.

\section{Feedback from the first stars} \label{sec:feedback}
In the previous section, we saw that the Pop.\ III IMF is likely top heavy, with a preponderance of massive stars. We know from the study of star formation in  the local Universe that massive stars, which are bright and short-lived, have a major impact on their environment: their radiation strongly heats their surroundings, dissociating H$_{2}$ and ionising atomic hydrogen; their stellar winds transfer substantial amounts of momentum and energy to the surrounding gas; and their deaths as supernovae further disrupt the gas while also enriching it with heavy elements (``metals''). Collectively, these processes are referred to as {\bf stellar feedback}. Given the importance of stellar feedback in present-day star formation, it is natural to ask whether it plays an equally important role in Pop.\ III star formation. To answer this, we discuss here the main forms of feedback from massive Pop.\ III stars and the role that they play in early star formation. 

\subsection{Stellar winds}
In present-day regions of massive star formation, stellar winds from O stars and Wolf-Rayet stars play a major role in regulating the star formation efficiency. However, these winds are driven by the radiation pressure exerted on the gas by the absorption of stellar photons in the large number of metal absorption lines present in the stellar envelope and circumstellar gas. As a result, the effectiveness of this form of wind driving is strongly metallicity dependent \citep{kud02}. As the metallicity decreases, the mass loss rate of the stars and the wind luminosity decrease substantially, and by the time we reach zero metallicity, wind driving is almost completely ineffective \citep{Vink2000}. 

One possible exception to this blanket statement would be if Pop.\ III stars were very rapid rotators. If they had rotational velocities of 50--80\% of the break-up velocity (i.e.\ the velocity at which a star is rotating so fast that it breaks apart), then the consequent rotational mixing inside the star could bring C, N and O from the core up to the surface, allowing the star to drive a strong wind \citep{ek08}. However, even in this case, this process only really becomes significant after the stars have already left the main sequence and started their He-burning phase. The general consensus in the field is therefore that stellar winds are probably not an important source of feedback from Pop.\ III star formation, although this is not yet completely settled.

{
\begin{figure}[t]
\hspace{0.3cm}
\setlength{\unitlength}{1cm}
\begin{picture}(15, 10.9)(0,0)
\put(-0.5,0){\includegraphics[width=0.55\textwidth]{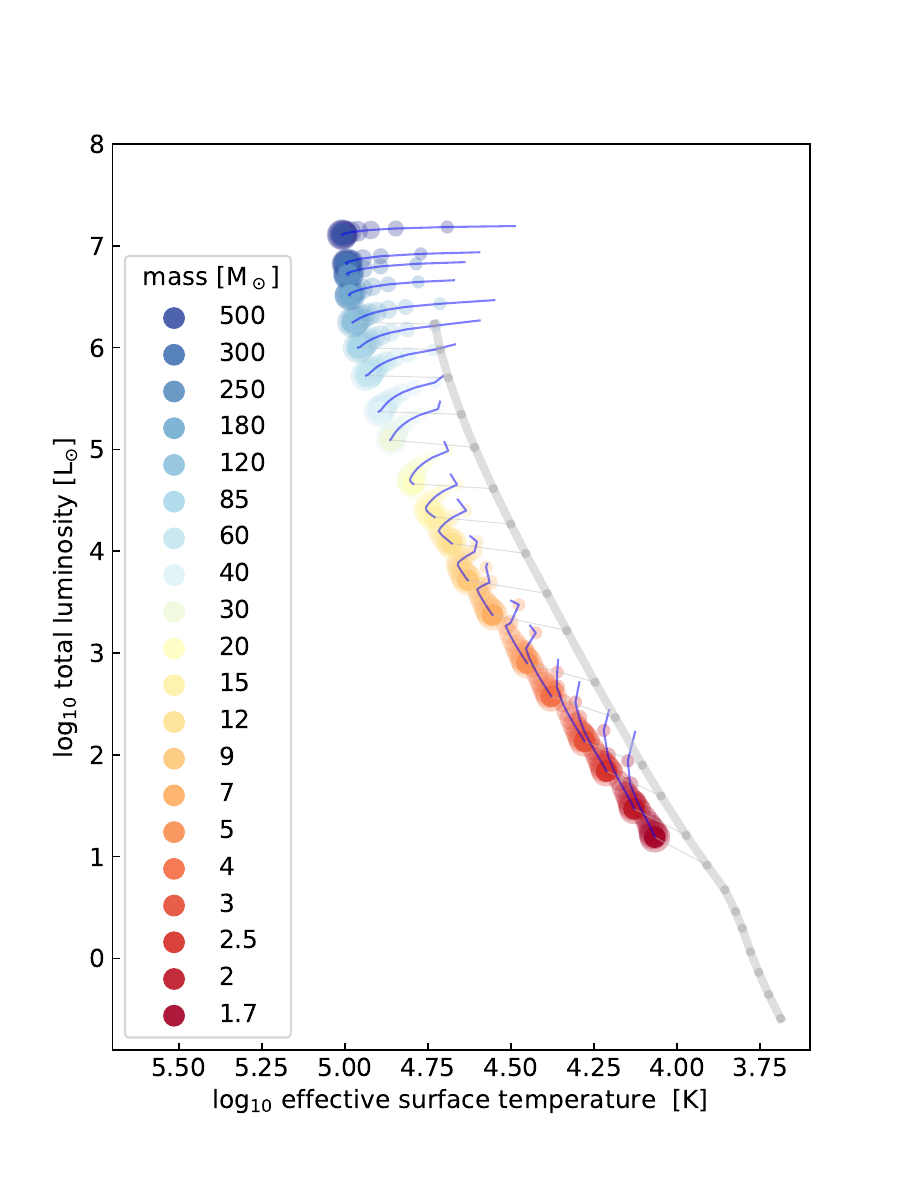}} 
\put( 7.9,3.79){\includegraphics[width=0.45\textwidth]{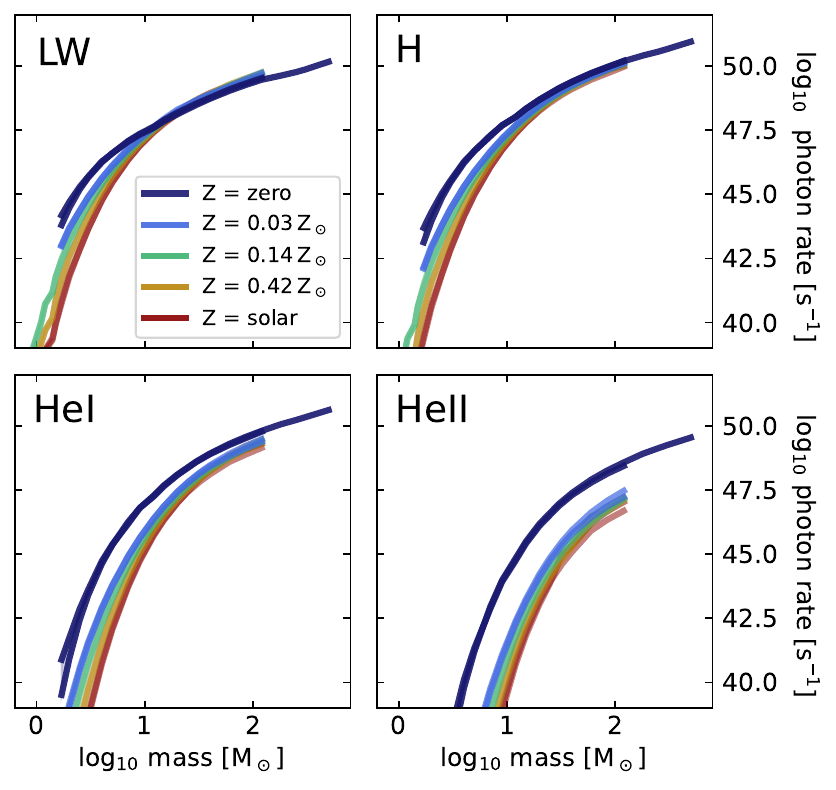}} 
\put( 8.0, 0.8){
\begin{minipage}[b]{0.44\textwidth}
\caption{
{\em Left:} Position of Pop.~III stars in the HRD across their MS and post-MS phases (as indicated by colour and different size symbols). We also show the zero-age main sequence loci of solar metallicity stars (gray). {\em Right:} Number of photons emitted per second averaged over the MS and post-MS lifetime in the LW-band and at H, He, and He$^+$ ionising wavelengths as function of mass and metallicity (colour coded). Modified version of a figure from \citet{kg23}.
\label{fig:HRD-flux}
}
\end{minipage}
}
\end{picture}
\vspace*{-0.5cm}
\end{figure}
}
\subsection{Radiative feedback}

Due to the lack of metals, Pop.~III stars are typically more compact and hence have higher surface temperatures than their present-day counterparts. Their location in the Hertzsprung-Russell diagram (HRD) is consequently shifted to the left (and also somewhat further up). To illustrate this, we plot at the left side of Figure~\ref{fig:HRD-flux} the $\log_{10} L - \log_{10} T$ position of  Pop.~III stars in the mass range $1.7 - 500\,$M$_{\odot}$ during their main sequence (MS) and post-MS evolution \citep[based on models by][computed with the Geneva stellar evolution code]{Murphy2021, Martinet2023}. For comparison, we depict the zero-age main sequence of solar metallicity stars \citep[from][]{Ekstrom2012} with masses $0.8 - 120\,$M$_{\odot}$. We also provide the lifetime-averaged fluxes for stars across the above mass range at the right side of Figure~\ref{fig:HRD-flux}. Specifically, we plot the number of {\bf Lyman-Werner (LW) photons} emitted per second (with energies in the range $11.2\,{\rm eV} \le h\nu < 13.6\,{\rm eV}$, see Section \ref{sec:LW}), as well as those that can ionise hydrogen (H: $h\nu \ge 13.6\,{\rm eV}$), neutral helium (HeI: $h\nu \ge 24.6\,{\rm eV}$), and singly ionised helium (HeII: $h\nu \ge 54.4\,{\rm eV}$) for a range of metallicities. It is clearly noticeable that metal-free Pop.~III stars have a much harder spectrum than their metal-enriched counterparts and are more capable of producing ionising photons. For further discussions and for tables of the lifetime-integrated values as a function of mass, we refer to the review by \cite{kg23}.

\subsubsection{Photoionisation}
\begin{figure}
\vspace{-0.3cm}
\includegraphics[width=0.65\textwidth]{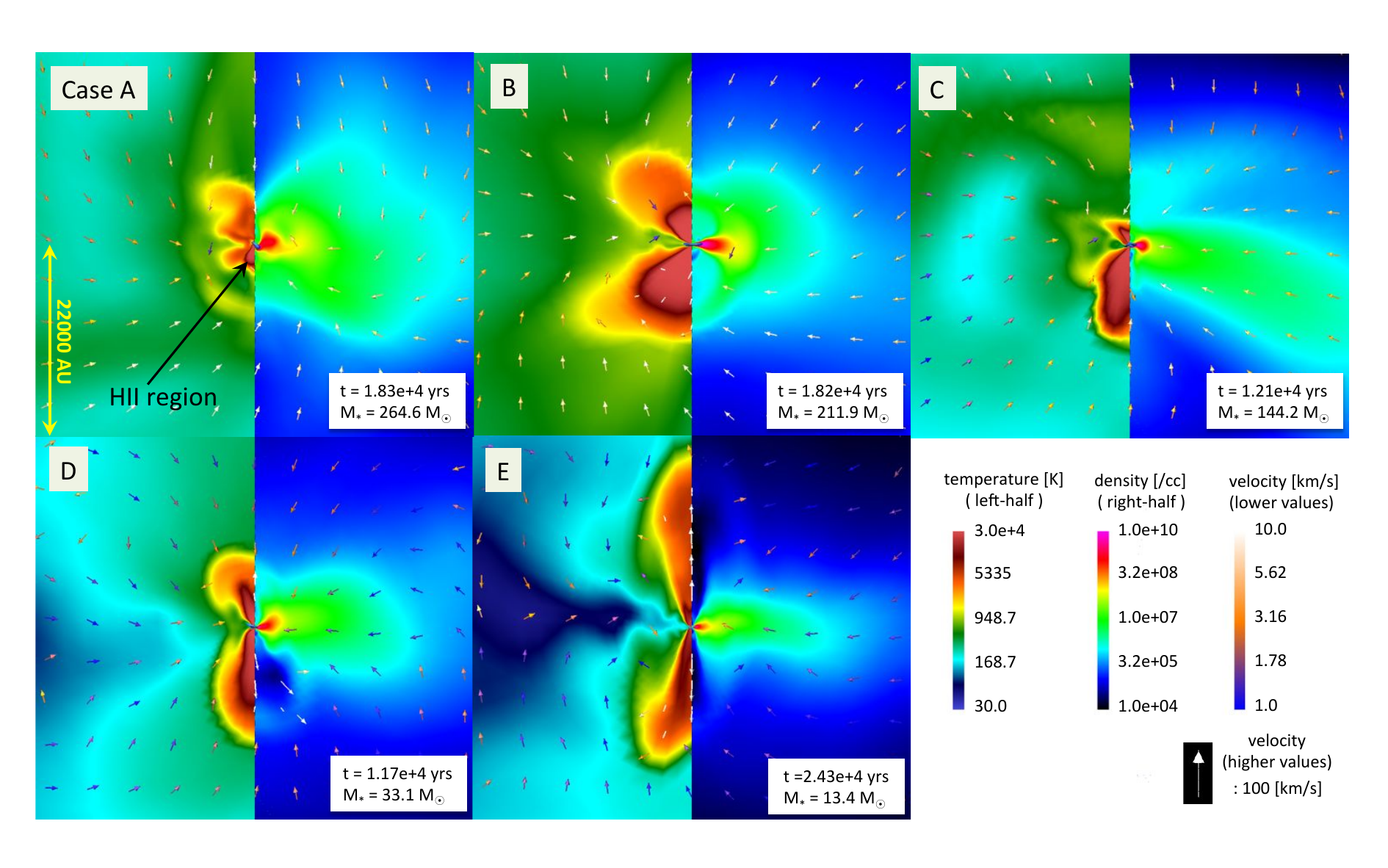}
\begin{minipage}[b]{0.30\textwidth}
\caption{Illustration of the characteristic bipolar morphology of the H{\sc ii} regions produced by massive Pop.\ III stars. Five examples are shown, taken from the simulations of \citet{hosokawa16}. Arrows indicate the velocity field, which is dominated by inflow in the neutral gas and by outflow in the ionised regions. The time for each simulation is indicated in the box, with $t = 0$ corresponding to the formation of the first protostar in the simulation. The box also indicates the mass of the central star at the moment depicted. Figure from \citet{hosokawa16}, reproduced with permission.
\vspace*{0.4cm}}
\label{fig:ion}
\end{minipage}
\vspace{-0.4cm}
\end{figure}

Massive Pop.\ III stars have large luminosities and high effective temperatures, and hence emit large numbers of photons capable of photoionising hydrogen \citep{Schaerer2002}. These photons are readily absorbed in the surrounding gas, leading to the formation of an H{\sc ii} region. The evolution of this H{\sc ii} region on scales much larger than the protostellar accretion disk is reasonably well understood. Once break-out occurs, it does so in the direction of the largest density gradient, which is typically perpendicular to the disk midplane, and leads to the formation of a rapidly expanding bipolar H{\sc ii} region (see Figure~\ref{fig:ion}). As it expands, it steadily reduces the supply of fresh neutral gas to the accretion disk. Over time, this leads to a reduction in the disk density. Simultaneously, gas is also being lost from the surface of the disk due to photoevaporation. These effects combine to eventually destroy the disk entirely after a few times $10^{4} \: {\rm yr}$, terminating both fragmentation and the accretion of gas onto existing protostars \citep[see e.g.][]{hosokawa16}. Following this, the H{\sc ii} region continues to expand in all directions. Photoionisation equilibrium is never reached owing to the steep density gradient in the halo and the fact that the sound speed of the ionised gas is higher than the escape velocity of the halo. The final outcome is therefore the loss of the majority of gas from the halo on a timescale of a Myr or less \citep[see e.g.][]{abel07}.

A more complicated question to answer is how the H{\sc ii} region first breaks out of the protostellar accretion disk, and when, if indeed it does. There are two main issues here. First, massive Pop.\ III stars that are accreting gas faster than a critical rate $\dot{M}_{\rm crit} \simeq 4 \times 10^{-3} \: {\rm yr}$ have inflated outer envelopes and hence low effective temperatures, $T_{\rm eff} \sim 6000$~K \citep{Hosokawa2009,Haemmerle2018}. These rapidly accreting massive stars therefore produce very few ionising photons and are hence unable to ionise their surroundings. Numerical simulations suggest that many massive Pop.\ III stars spend at least part of their early lives in this high accretion state, but the total duration of this phase is unclear \citep{ham20}.

Second, even after the Pop.\ III star has thermally relaxed to the main sequence and started producing ionising photons, the H{\sc ii} region that it produces may remain trapped in the dense gas close to the star. This is the case as long as the Str\"omgren radius, obtained from detailed balance between photoionisation and recombination, is much smaller than the local disk scale-height. It can occur if the local gas density is very high. In this scenario, although the ionised gas is over-pressured compared to its surroundings, it is located so close to the star that it is gravitationally bound to it, preventing pressure-driven expansion of the H{\sc ii} region. Numerical simulations of Pop.\ III H{\sc ii} region growth that do not resolve the disk structure in the vertical direction cannot determine whether or not H{\sc ii} region trapping occurs, but recent extremely high resolution simulations suggest that it may occur frequently \citep{jaura22}. However, these simulations do not include an important physical effect -- radiation pressure due to Lyman-$\alpha$ photon scattering -- that may dominate the early dynamical evolution of the H{\sc ii} region and allow it to escape from the disk. The final resolution to this issue therefore remains uncertain.

\subsubsection{Photodissociation}
\label{sec:LW}
Stars that are massive enough to emit copious amounts of ionising radiation will also produce many photons capable of photodissociating H$_{2}$. These have energies in the Lyman-Werner band ($11.2\,{\rm eV} \le h\nu < 13.6\,{\rm eV}$) and can penetrate into the gas ahead of the ionisation front that bounds the H{\sc ii} region. At very high gas densities the separation between the ionisation front and the edge of the photodissociation region (PDR) is very small, owing to the absorption of a substantial number of these photons in the dense layer of atomic gas that builds up in the PDR. Therefore, LW photons have a very limited impact on the smallest scales in Pop.\ III star-forming systems.
In regions with density below $n \sim 10^{8} \: {\rm cm^{-3}}$, however, substantial separation between the H{\sc ii} region and the outer edge of the PDR starts to occur. In this lower density regime, Pop.\ III stars efficiently destroy H$_{2}$, heating the gas up to $T \sim 6000$~K in the process. This helps to reduce the supply of cold gas to the protostellar accretion disk, although simulation suggest that the overall effect is less significant than that of the ionising radiation 
\citep[see e.g.][]{sugimura23}. Finally, a significant fraction of the photons in the Lyman-Werner energy range relevant for H$_{2}$ photodissociation are able to escape from the halo entirely, with consequences that are discussed in the next section.

\subsubsection{Impact on the intergalactic medium}
In addition to the small-scale impact discussed above, radiative feedback from massive Pop.\ III stars can also act on much larger scales in the intergalactic medium (IGM). Short-wavelength radiation that escapes into the IGM can produce large ionised regions surrounding each star-forming halo, thanks to the low density of the IGM at these epochs. This can act to suppress star formation in nearby halos. Models show that the effect is strongest for systems in an early evolutionary stage, with peak densities below $100 \: {\rm cm^{-3}}$. Substantial amounts of the gas can then be ionised by the external radiation field on a timescale shorter than the free-fall collapse time \citep{whalen08}. On the other hand, systems with much higher peak densities tend to evolve more rapidly than they are ionised, resulting in a negligible impact on the progression of star formation. This is corroborated by the short duration of Pop.\ III star formation in typical low-mass halos and the rapid transition to Pop.\ II star formation that occurs following the explosion of the first supernovae. For this reason, photoionisation by Pop.\ III-dominated sources is unlikely to be the dominant form of feedback on large scales in the high redshift Universe.

In contrast, LW photons can easily escape from low mass protogalaxies. Since the IGM is largely transparent to these photons, they can travel great distances before being absorbed. The onset of Pop.\ III star formation in a given volume of space is therefore quickly followed by the growth of a large-scale Lyman-Werner radiation background in that volume \citep{har00}. The corresponding increased H$_2$ photodissociation rate in newly-assembling protogalaxies, reduces the ability of the gas to cool and thereby increases the critical halo mass scale $M_{\rm crit}$ above which we expect Pop.\ III star formation to occur. Unfortunately, the strength of the Lyman-Werner background required to significantly affect $M_{\rm crit}$ remains uncertain. Models that assume that  H$_{2}$ is optically thin to  LW photons generally agree that the LW background starts to significant alter $M_{\rm crit}$ once its mean specific intensity exceeds a critical value $J_{\rm crit} \sim 10^{-24} \: {\rm erg \, s^{-1} \, cm^{-2} \, Hz^{-1} \, sr^{-1}}$, a value that is reached relatively early after the onset of Pop.\ III star formation \citep{har00}. However, in reality H$_{2}$ molecules can {\bf self-shield}: molecular hydrogen located deeper in a halo can be protected from photodissociation by the absorption of LW photons by H$_{2}$ molecules in the outskirts of the halo. Some numerical models that attempt to account for this self-shielding find that it has little effect on $J_{\rm crit}$, while others find that it has a profound effect, increasing it by a factor of 1000 or more \citep[see e.g.\ the comparison of models in][]{kg23}, rendering LW feedback of little importance on large scales. The origin of this discrepancy remains unclear.

\subsection{Supernovae}
\label{sec:SNe}
\subsubsection{Varieties of Population III supernova}
The final form of stellar feedback that we have to consider here are the supernova explosions that accompany the death of many massive stars. In the absence of rotation and mass-loss, the final fate of a massive Pop.\ III star depends only on its main sequence mass. Several different outcomes are possible \citep{heger03}:
\begin{enumerate} 
	\item[i] Stars with masses between 10 and 40~$\rm M_{\odot}$ explode as conventional {\bf core-collapse supernovae} (CCSN).    
    \item[ii] Stars with masses between 40 and $\sim 70 \: \rm M_{\odot}$ collapse directly to form a black hole without an explosion.
    \item[iii] Stars with masses between $\sim 70$ and 140~M$_{\odot}$ undergo the pulsational pair instability, before eventually collapsing to form a black hole.
    \item[iv] Stars with masses between 140 and 260~$M_{\odot}$ explode as {\bf pair instability supernovae} (PISN).
    \item[v] Stars with $M > 260 \: {\rm M_{\odot}}$ directly collapse into black holes, again without an explosion. 
\end{enumerate}

Population~III core-collapse supernovae proceed in the same fashion as core-collapse supernovae produced by higher metallicity stars: the formation of a degenerate iron core that is unable to support itself against gravity is followed by a collapse that triggers a neutrino-powered explosion. For stars with masses below 25~M$_{\odot}$, this explosion has an energy of around $10^{51}$~erg and leaves behind a neutron star remnant, while for stars with masses between 25~M$_{\odot}$ and 40~M$_{\odot}$, a black hole results. In this case, fallback of matter and energy into the black hole can lead to a so-called faint supernova with a low yield of elements heavier than C or O and an energy significantly below the canonical value. 

Above $70 \: {\rm M_{\odot}}$, a new physical effect becomes important that can lead to a fundamentally different type of supernova explosion. When these highly massive Pop.\ III stars leave the main sequence, they contract, raise their central temperatures, and begin to fuse helium. The central temperature becomes so high that they start to produce a substantial number of photons with energies above the rest mass energy of an electron-positron pair. These photons can be removed by pair creation, i.e.\ the spontaneous creation of an electron-positron pair. This triggers an instability: the removal of high energy photons decreases the radiation pressure, which leads to contraction of the star and an increase of the central temperature with an even higher pair production rate. The result is a run-away contraction of the core. For stars with masses between $\sim 70 \, {\rm M_{\odot}}$ and $\sim 140 \, {\rm M_{\odot}}$ the process halts once rapid O and Si burning sets in. The instability becomes pulsational with the star undergoing a series of contractions, nuclear flashes and re-expansions. The energy released by this pulsational pair instability ranges from $\ll 10^{51} \, {\rm erg}$ to a few times $10^{51} \, {\rm erg}$, and the ejected mass also spans a large range of values. The pulsations eventually fade away once the star has lost enough mass to bring it below the lower mass threshold. It now evolves similarly to the lower mass stars. Above $140 \: {\rm M_{\odot}}$, the same pair creation instability occurs, but in this case so much energy is released in the initial nuclear flash that it completely destroys the star, resulting in a pair instability supernova explosion. These explosions have energies ranging from a few times $10^{51}$~erg to $10^{53}$~erg, substantially higher than the energy of a CCSN. In addition, because they leave no remnant, these explosions typically produce very large quantities of metals.

Accounting for rotation has two main effects on this picture \citep{Maeder2000}. First, rotational-driven mixing can lower the mass required for a PISN, potentially to a value as low as $100 \: {\rm M_{\odot}}$. Second, rapidly rotating Pop.\ III stars can potentially explode as jet-driven supernovae \citep{grimmett21}, producing {\bf hypernova} explosions with energies of $10^{52} \: {\rm erg}$. However, whether Pop.\ III stars rotate rapidly enough for either of these effects to become important remains highly unclear.

\subsubsection{Feedback from Population III supernovae}
The energy released by even a single CCSN is higher than the gravitational binding energy of the gas in a $10^{6} \: {\rm M_{\odot}}$ halo. Therefore, 
once Pop.\ III stars explode, they rapidly clear out the majority of the gas from the first generation of star-forming halos. These halos subsequently re-accrete gas from their surroundings, and hence can eventually begin forming stars again, but the time delay between the explosion of the Pop.\ III supernovae and the renewed onset of star formation -- a quantity known as the {\bf recovery time} -- can be considerable, ranging from $10 - 100\,$Myr \citep[see e.g.][]{magg22}. During this period, the halo typically has grown considerably in size and mass. In addition, it has been enriched by heavy elements produced by the previous supernovae. This defines the environment where the early generations of Pop.~II stars form. 

To better understand this early enrichment history, it is useful to look at the metallicity produced by a single Pop.\ III supernova. If we assume perfect mixing within the enriched region, then the change in metallicity is
\begin{equation}
\Delta {\rm Z} = \frac{M_{\rm metal}}{M_{\rm region}}, 
\end{equation}
where $M_{\rm metal}$ is the total mass of metals produced by the supernova (often referred to as its {\bf yield}), and $M_{\rm region}$ is the total
mass of the region enriched by the supernova. A lower limit on the latter comes from the mass of gas swept up by the expanding supernova remnant. This depends on the supernova energy, the density of the region in which the supernova explodes and the mass of gas available within the halo. Typical values are $10^{4}$--$10^{5} \: {\rm M_{\odot}}$ for a CCSN and more than $10^{6} \: {\rm M_{\odot}}$ for a high-energy PISN. However, once one accounts for the fact that PISNe produce a much larger yield of metals than CCSNe, the resulting maximum metallicity is actually fairly similar in both cases with ${\rm Z_{\rm max}} \sim 10^{-3} \: {\rm Z_{\odot}}$ \citep{Magg2020}. Stars forming from this metal-enriched gas can end up with lower metallicities if additional mixing with fresh metal-free gas occurs prior to the formation of the second generation stars, but it is hard to envisage mechanisms that will lead to the formation of true second generation stars with metallicities ${\rm Z} \gg {\rm Z_{\rm max}}$. Therefore, when looking for the signatures of Pop.\ III supernovae, we should look at the most metal-poor stars in our Galaxy, a point we will return to in Section~\ref{sec:arch}.

\section{Observing the first stars}\label{sec:obs}
\subsection{Direct detection of Pop.\ III stars}
One of the unusual features of the study of the first generation of stars is the almost complete absence of observational data. This stands in stark contrast to the situation in present-day star formation, where we are almost overwhelmed with information. Direct observations of the sites of Pop.\ III star formation would provide important constraints on our theoretical models. Unfortunately, as we explore in this section, observing the first stars in situ is extremely challenging.

The simplest way in which we might try to observe the first stars or star clusters is to directly image them with an instrument such as NIRCAM or MIRI on the James Webb Space Telescope (JWST). However, the fluxes of individual stars and stellar clusters are far below the limit we can practically reach with JWST. For example, \citet{schauer20} show that for a $1000 \: {\rm M_{\odot}}$ cluster of Pop.\ III stars, all of which are assumed to be very massive, the observed flux ranges from $10^{-2}$~nJy at $z = 10$ to $2 \times 10^{-3}$~nJy at $z = 30$. These numbers assume a very young cluster and account for the nebular emission as well as the direct continuum emission from the star. To put these numbers into context, for a $10^{4} \: {\rm s}$ exposure, NIRCAM has a point source sensitivity of around $10$~nJy in its most sensitive filters.\footnote{See \href{https://jwst-docs.stsci.edu/jwst-near-infrared-camera/nircam-performance/nircam-sensitivity\#gsc.tab=0}{https://jwst-docs.stsci.edu/jwst-near-infrared-camera/nircam-performance/nircam-sensitivity\#gsc.tab=0}.}. Note that the point source sensitivity of MIRI is an order of magnitude worse at the shortest wavelengths that MIRI is sensitive to and even worse at longer wavelengths, so stars that are too faint to detect with NIRCAM will almost certainly also be too faint to detect with MIRI. In the simple case in which we are dominated by Gaussian noise, so that the signal-to-noise scales with time as $t^{1/2}$, this implies that JWST would need to observe the same patch of sky for approximately $10^{10} \: {\rm s} \simeq 300 \: {\rm yr}$ to observe these stars! 

In order for us to be able to detect a cluster of Pop.\ III stars, we either need to find a system that is much more massive than the clusters considered above, or its brightness in the observer's frame must be significantly boosted by gravitational lensing. The first of these possibilities appears unlikely, at least within the context of $\Lambda$CDM. As we discussed in Section~\ref{sec:where}, the first Pop.\ III star formation occurs in dark matter halos with total masses of $1 - \mathrm{few} \times 10^6\: {\mathrm M_{\odot}}$. The gas masses associated with these halos are a factor of $\Omega_{\rm b} / \Omega_{\rm m}$ smaller, and are hence several $10^{5} \: {\rm M_{\odot}}$. To directly detect a young Pop.\ III cluster without lensing, we would need it to have a mass of around $10^{5} \: {\rm M_{\odot}}$ (assuming a top-heavy IMF), implying a required star formation efficiency of between 20 and 50\%. This is much higher than the observed efficiency of star formation in local star-forming regions, which on $50 - 100\,$pc scales is typically a few percent \citep[see e.g.][]{sun23}. It seems difficult to bring about in metal-free gas, given the relatively long cooling time of this gas.\footnote{Note that in metal-enriched systems with much shorter cooling times, it may be possible to produce massive, dense clouds of gas quickly enough to achieve very high star formation efficiencies, even in the presence of stellar feedback \citep[see e.g.][]{Dekel2023, Polak2024}.} In order to produce a co-eval cluster with a mass of $20 - 50$\% of the initial gas mass, star formation would need to be delayed until after this much gas has cooled, but not suppressed entirely within the halo. To the best of our knowledge, behaviour of this kind is not found in any simulations of Pop.\ III star formation with sufficient resolution to model star formation in the earliest generation of star-forming dark matter halos.

In order to build up sufficiently massive clusters of Pop.\ III stars without requiring an extremely high star formation efficiency, it is therefore necessary to form them in dark matter halos with masses $\gg 10^{6} \: {\rm M_{\odot}}$. However, it is not easy to arrange for the gas in such systems to remain completely metal-free: typically, by the time we are dealing with halos massive enough to be able to accumulate enough dense gas to form $\sim 10^{5} \: {\rm M_{\odot}}$ of Pop.\ III stars, one or more of their progenitor systems will already have formed stars and hence will already have become metal-enriched. 

An additional difficulty is that even if it is possible in some cases to assemble a sufficiently massive cloud of dense, cool, metal-free gas, it is not clear that this will necessarily result in the formation of a cluster of UV-bright stars. In these conditions, stellar mergers will be common and may lead to the run-away build up of a supermassive star \citep[see e.g.][]{reinoso23}, with $M > 10^{4} \: {\rm M_{\odot}}$. A supermassive star formed in this way would have a very high luminosity, but would also have a large radius and hence a low effective temperature, and hence would likely be far too faint in the rest-frame ultraviolet to be visible with NIRCAM.

The other possibility -- boosting the observed brightness of the Pop.\ III stars through lensing -- initially seems more promising. Although the required magnifications are large ($\mu > 1000$), such values can be produced by massive foreground lenses (e.g.\ galaxy clusters) in regions on the sky sufficiently close to a lensing caustic. Indeed, Hubble Space Telescope (HST) and JWST observations of lensing clusters have already found at least one high redshift object that is magnified by a factor of well over 1000, although it is still not clear whether this star (WHL~0137-LS, nicknamed ``Earendel'') is a Pop.\ III star or an early Pop.\ II star \citep{welch22}.

The main limitation when it comes to using high magnification lensing to search for Pop.\ III stars is simply the very small fraction of the sky that is actually magnified by a factor $\mu > 1000$. Because of this, observations made at high magnification only probe small volumes of the high redshift Universe. Therefore, it is highly likely that for any given foreground lens, there simply will not be any Pop.\ III star formation occurring in the volumes probed by the highest magnification factors. Even if one makes optimistic assumptions regarding the rate at which Pop.\ III stars are forming in the early Universe, the best current estimates \citep{zack24} suggest that the probability of detecting any highly magnified Pop.\ III stars in observations of a single lensing cluster is $\ll 10$\%. Surveys of large numbers of lensing clusters are therefore required, and even then we may in the end still be unlucky and fail to find any Pop.\ III stars.

\subsection{Direct detection of Pop.\ III supernovae}
As well as trying to detect massive Pop.\ III stars during their lives, we can also attempt to detect the supernova explosions that they produce at the end of their lives. The high luminosities of Pop.\ III core-collapse and pair instability SNe compared to main sequence Pop.\ III stars makes them much easier to detect without the need for high magnification gravitational lensing. For example, \citet{whalen13a,whalen13b} show that Pop.\ III CCSNe should be visible for tens to hundreds of days, depending on the progenitor mass, while PISNe should be visible for hundreds to thousands of days.

However, because Pop.\ III supernovae remain bright for only a short time, the number that we expect to find on the sky at any given moment is highly limited. This number is sensitive to the cosmic evolution of the Pop.\ III star formation rate density and the IMF of Pop.\ III stars. In the most optimistic models, we expect to detect $\sim 1$ Pop.\ III PISN within the whole area covered by existing or currently ongoing deep JWST surveys \citep{Venditti24}; in more pessimistic models, the corresponding number is only $\sim 0.05$ (i.e.\ our chances of detecting even a single Pop.\ III PISN could be as small as a few percent). The reason that these numbers are so small, despite JWST's extremely high infrared sensitivity, is the small field of view of the telescope, and so a greater number of discoveries may actual come from the Nancy Grace Roman Space Telescope, which has a worse sensitivity but a much larger field of view \citep{moriya19}.
The benefits of high sky coverage have also motivated an investigation into whether Pop.\ III PISN will be detectable by the Simonyi Survey Telescope\footnote{Previously known as the Large Synoptic Survey Telescope, LSST.} at the Vera C.\ Rubin Observatory, with the conclusion being that this telescope may detect some lensed Pop.\ III PISN in its Deep Drilling fields, depending on just how deep these are, but will likely not detect any during its baseline survey 
\citep{Rydberg2020}. For Pop.\ III CCSNe, the number likely to be found by JWST is at most an order of magnitude larger, but these supernovae will be too faint to detect with Roman and the chances of detecting them with the 
Simonyi telescope are also small. Overall, although no Pop.\ III supernova have as yet been detected, it is plausible that some will be within the next few years. 

\subsection{Stellar archaeology}
\label{sec:arch}

As well as trying to observe the first stars and supernovae in situ, we can also study their effects indirectly by looking at the imprint they leave on later generations of star formation. This is the field of {\bf stellar archaeology}. It is possible because the lifetime of a star is a strong function of its mass. While high-mass stars live for only a few to a few tens of Myr, lower mass stars remain on the main sequence for much longer times. For example, our own Sun has a main sequence lifetime of around $10^{10}$~yr. Stars slightly less massive than the Sun, below $\sim 0.8 \: {\rm M_{\odot}}$,  have main sequence lifetimes that exceed the age of the Universe, and hence any such stars that formed at high redshift will remain available for observation today. If we find examples of these stars, we can measure the relative abundances of different elements present in their atmospheres and put together a picture of the elemental composition of the gas at the time that the stars formed. This in turn allows us to put constraints on the properties of the supernovae responsible for producing these elements.

Directly age dating low mass main sequence stars is very difficult, and so to identify stars likely to have been formed at high redshift, we generally rely on their metallicity as a proxy, i.e.\ we assume that lower metallicity stars are, on average, older than higher metallicity stars. Specifically, if we restrict our attention to {\bf extremely metal-poor stars} (EMP) with metallicities\footnote{We use the standard notation for specifying elemental abundances: $[{\rm X} / {\rm H}] = \log_{10} (N_{\rm X} / N_{\rm H}) - \log_{10}(N_{\rm X, \odot} / N_{\rm H, \odot})$, where $N_{\rm X}, N_{\rm H}$ are the fractional abundances by number of element X and hydrogen, respectively, and $N_{\rm X, \odot}, N_{\rm H, \odot}$ are the corresponding values in solar metallicity gas.} $[{\rm Fe}/{\rm H}] < -3$,  it is likely that the vast majority were indeed formed at high redshift. These stars are of great interest for the study of Pop.\ III because their metallicities are low enough that they could potentially have been formed from gas enriched with the metals produced in only a single Pop.\ III supernova \citep{Salvadori2019, Magg2020, Vanni2023}. They therefore offer the cleanest insight into the properties of massive Pop.\ III stars. 
Comprehensive discussions of the different ways in which stellar archaeology can be used to probe the properties of early stellar populations can be found in the reviews by \citet{beers05} and \citet{frebel15}. Here, we focus on three important results relevant to Pop.\ III.

Concerning the low-mass end of the Pop.\ III IMF, we have yet to find any truly metal-free stars. Although the number of such stars is expected to be small compared to the number of metal-enriched stars, our failure to find any examples allows us to already rule out models in which Pop.\ III stars form with the same IMF as present-day stars, as this would result in the formation of too many Pop.\ III stars with lifetimes longer than the age of the Universe. Therefore, either the minimum Pop.\ III stellar mass must be larger than $0.8 \: {\rm M_{\odot}}$ (in which case all of the stars would already have evolved off the main sequence by the present day) or the low-mass portion of the Pop.\ III IMF must be much flatter than the corresponding part of the present-day IMF \citep{magg19,rossi21}. Future observations will  improve these constraints.

Furthermore, the majority of the EMP stars that have been discovered are what is known as {\bf carbon-enhanced metal poor} (CEMP) stars, i.e.\ they have a carbon to iron ratio that is much higher than the ratio found in the Sun. Many of these stars also show evidence for s-process enrichment, indicating that they have been enriched by elements produced in an asymptotic giant branch (AGB) star, likely a binary companion of the star in question, and for these stars, this AGB enrichment may also explain the elevated C/Fe ratio. However, there is also a subset of CEMP stars, known as CEMP-no stars, that show no s-process enrichment. In these stars, the elevated C/Fe ratio must reflect the abundance pattern of the material out of which the star formed, and hence directly represents the yields of the supernova enriching the gas. A high C/Fe ratio is indicative of the production of very little iron by the supernova, and so the preponderance of CEMP-no stars at very low metallicity may indicate that faint, low energy supernovae play an important role in enriching this gas with typical progenitor masses in the range $20 - 40\,$M$_{\odot}$ \citep[see e.g.][]{yoon19}.

At the high-mass end of the IMF, models of pair instability supernovae predict that they should produce elemental yields with a very pronounced ``odd-even'' effect, i.e.\ elements with even atomic numbers (e.g.\ C, O, Ne) should be produced in much larger quantities than elements with odd atomic numbers (e.g.\ N, F, Na). Very few examples of such an odd-even pattern have actually been found \citep[see e.g.][and references therein]{skuladottir24}. This is important, as it demonstrates that most of the supernovae occurring at high redshift must have been some kind of core-collapse supernova, with PISN explosions being relatively rare events \citep[see also][]{Koutsouridou2024}. This in turn directly constrains the high mass end of the Pop.\ III IMF, ruling out models that only produce stars with masses $M \gg 50 \: {\rm M_{\odot}}$. 

\subsection{Gravitational waves}

The detection of {\bf gravitational waves} (GW) through advanced interferometers like LIGO, Virgo, and KAGRA, and upcoming missions such as LISA and DECIGO and eventually the Einstein Telescope (ET), provides an alternative avenue to study the high-redshift Universe. Current signals primarily arise from mergers of compact binaries, including black holes and neutron stars, and their analysis offers insights into stellar evolution, cosmic star formation, and binary dynamics. 
Using gravitational wave data, we can infer properties of the merging objects such as their mass, spin, and merger rates. Many of these mergers will involve the remnants of Pop.\ II stars, but some will involve the remnants of Pop.\ III stars. Observations of gravitational waves can therefore help to constrain the IMF and the formation pathways of Pop.\ III stars. However, interpreting these signals is challenging due to complexities in binary evolution, cluster dynamics, and the rarity of Pop.\ III star remnants compared to metal-enriched systems \citep[for a review, see e.g.][]{Mandel2022}. 	

The mass spectrum of merging black holes from the combined data of the first three observing runs of LIGO, Virgo, and KAGRA \citep{Abbott2023} shows peaks around $\sim 10\,$M$_\odot$ and $30–35\,$M$_\odot$. Although most objects detected have masses below $45\,$M$_\odot$, the distribution extends beyond $\sim 65\,$M$_\odot$ and reaches into the so-called ``pair-instability gap'' (at least for metal-rich stars, see also Section \ref{sec:SNe}). This is in principle consistent with a Pop.\ III origin. However, the current event rates are much too high for this to be the primary explanation \citep[e.g.][]{Hartwig2016, Tanikawa2022, Santoliquido2023}. It is more likely that dynamical formation pathways involving stellar collisions and mergers are involved \citep[e.g.][]{Mapelli2016}. This is further strengthened by the fact that a significant fraction of observed GW events indicate spin misalignment, as is expected from dynamical interactions in dense stellar clusters. 

In summary, while current GW detections have not yet provided direct evidence of Pop.\ III stellar remnants, they provide important constraints on theoretical models regarding the mass spectrum, merger rates, and formation environments of stellar remnants. These findings suggest that the majority of observed GWs originate from metal-enriched stars, while Pop.\ III stars likely remain hidden, contributing primarily at higher redshifts and requiring considerably larger sample sizes or next-generation detectors, such as LISA or the Einstein Telescope, for their potential identification.

\section{Summary}\label{summary}
 The first stars form at redshifts $z < 30$ in dark matter halos with masses of around $10^{6} \: {\rm M_{\odot}}$. This critical halo mass scale is set by the requirement that the gas be able to form enough H$_{2}$ to cool efficiently within a dynamical time. Rapid dynamical heating (e.g.\ by frequent mergers) or a high streaming velocity of the baryons relative to the dark matter can significantly increase this minimum mass scale.

 The thermal evolution of Pop.\ III star-forming gas is governed on almost all scales by H$_{2}$. Cooling remains effective over many orders of magnitude in density, allowing the gas to gravitationally collapse and form stars. The initial contraction and formation of a first protostar is rapidly followed by the build-up of a gravitationally-unstable protostellar accretion disk. Fragmentation of this disk then results in the formation of a  dense cluster of metal-free protostars. Because the metal-free gas in these systems is much hotter than the gas in typical local star-forming regions, the mass inflow rate is also much higher. The stars in this dense cluster therefore gain mass rapidly. The resulting IMF is uncertain, as no completely comprehensive model for it yet exists, but all indications are that it is much flatter than the present-day IMF, with the majority of the stellar mass accounted for by stars much above one solar mass. 
 
 Feedback from Pop.\ III stars plays an important role in regulating further star formation. Within the dark matter halo, photoionisation dominates (possibly assisted by Lyman-$\alpha$ radiation pressure at early times) and begins to quench star formation after a few times $10^{4} \: {\rm yr}$. Once the  first supernovae occur, roughly two million years after the onset of star formation, all remaining gas gets removes from the host halo and dispersed into the intergalactic gas. 
 On much larger scales, the build-up of an extragalactic Lyman-Werner background starts to suppress Pop.\ III star formation in low-mass dark matter halos, while chemical enrichment of the gas by SNe drives the transition from Pop.\ III to Pop.\ II star formation.

 On the observational side, direct detection of individual Pop.\ III stars or star clusters appears to be highly unlikely in the forseeable future, even with the aid of high magnification gravitational lensing. On the other hand, detection of at least a few Pop.\ III SNe within the next decade appears to be promising. At  present, however, the strongest observational constraints on Pop.\ III stars are the indirect ones that we can derive from stellar archaeology. Our failure to find any surviving Pop.\ III stars constrains the low-mass end of the Pop.\ III IMF, while the almost complete absence in metal-poor stars of the characteristic odd-even abundance pattern produced by pair-instability supernovae constrains the Pop.\ III IMF at high stellar masses. Current observations point towards core collapse supernovae with progenitor masses in the range $20 - 40\,\mathrm{M}_\odot$ as being the most likely cause of early enrichment and the trigger for the transition to Pop.\ II star formation. These constraints will continue to improve as the size of our sample of extremely metal-poor stars continues to increase. We note that future gravitational wave detections will also help to constrain the parameters of Pop.\ III stars.

\begin{ack}[Acknowledgments]

The authors acknowledge financial support from the European Research Council via the ERC Synergy Grant ``ECOGAL'' (project ID 855130), from the German Excellence Strategy via the Heidelberg Cluster of Excellence (EXC 2181 - 390900948) ``STRUCTURES'', and from the German Ministry for Economic Affairs and Climate Action in project ``MAINN'' (funding ID 50OO2206). They are also grateful for computing resources provided by the Ministry of Science, Research and the Arts (MWK) of the State of Baden-W\"{u}rttemberg through bwHPC and the German Science Foundation (DFG) through grants INST 35/1134-1 FUGG and 35/1597-1 FUGG, and also for data storage at SDS@hd funded through grants INST 35/1314-1 FUGG and INST 35/1503-1 FUGG. RSK also thanks the Harvard Radcliffe Institute for Advanced Studies and the Harvard-Smithsonian Center for Astrophysics for their hospitality during his sabbatical, and the 2024/25 Class of Radcliffe Fellows for highly interesting and stimulating discussions.
\end{ack}

\seealso{Recent reviews of Population III star formation that cover many of these topics in greater detail can be found in \citet{ham20} and \citet{kg23}.}

\bibliographystyle{Harvard}
\bibliography{reference}

\end{document}